\documentstyle[12pt,moriond,psfig]{article}
\def \baas#1#2  {{\em Bull.\ American Astron.\ Soc.\/} {\bf #1}, {#2}}
\def\ltsim{~\rlap{$<$}{\lower 1.0ex\hbox{$\sim$}}~}
\def\gtsim{~\rlap{$>$}{\lower 1.0ex\hbox{$\sim$}}~}
\newcommand{\HI}{\mbox {\sc H\thinspace{i}}}
\newcommand{\HII}{\mbox {\sc H\thinspace{ii}}}
\newcommand{\Hline}[1]{\mbox{H{\footnotesize {#1}}}}
\newcommand{\Halpha}{\Hline{\mbox{$\alpha$}}}
\newcommand{\kms}{\mbox{km\thinspace s$^{-1}$}}
\newcommand{\Mlbstar}{\mbox{$({\cal M}/L_B)_\star$}}
\newcommand{\Msun}{\mbox{${\cal M}_\odot$}}
\begin{document}
\heading{STARBURSTS, DARK MATTER, AND \\
THE EVOLUTION OF DWARF GALAXIES}

\author{Gerhardt R. Meurer}{The Johns Hopkins University}{Baltimore MD,
USA}

\begin{moriondabstract}
Optical and \HI\ imaging of gas rich dwarfs, both dwarf irregulars (dI)
and blue compact dwarfs (BCD), reveals important clues on how dwarf
galaxies evolve and their star formation is regulated.  Both types
usually show evidence for stellar and gaseous disks.  However, their
total mass is dominated by dark matter.  Gas rich dwarfs form with a
range of disk structural properties.  These have been arbitrarily
separated them into two classes on the basis of central surface
brightness.  Dwarfs with $\mu_{0}(B) \ltsim 22\, {\rm mag\,
arcsec^{-2}}$ are usually classified as BCDs, while those faintwards of
this limit are usually classified as dIs.  Both classes experience
bursts of star formation, but with an absolute intensity correlated with
the disk surface brightness.  Even in BCDs the bursts typically
represent only a modest\ltsim{1} mag enhancement to the $B$ luminosity
of the disk.  While starbursts are observed to power significant
galactic winds, the fractional ISM loss remains modest.  Dark matter
halos play an important role in determining dwarf galaxy morphology by
setting the equilibrium surface brightness of the disk.
\end{moriondabstract}

\section{Introduction}

Dwarf galaxies come in a variety of morphologies including: (1) dwarf
elliptical (dE) - defined by smooth elliptical isophotes, and which
invariably have a red color; (2) dwarf irregular (dI) - characterized
by an irregular structure, blue colors and \HII\ regions scattered
haphazardly over the optical face of the galaxy; and (3) blue compact
dwarf (BCD) \cite{tm81} - also an irregular morphology, differing from
dI in that the \HII\ emission is usually highly concentrated into one or
two high intensity patches near the center.  BCD galaxies frequently are
given other classifications such as Amorphous \cite{sb79,gh87}
(sometimes prefaced with the adjectives ``dwarf'' or ``blue''), or \HII\
galaxies \cite{tmmmc91}.  However, the properties of {\em dwarf\/} ($M_B
> -18$ mag) galaxies having these different monickers are virtually
identical \cite{mmhs97,mmh98}, indicating that they represent the same
physical phenomenon, hence I will refer to them all as BCDs.

Despite their morphological differences dIs, BCDs, and dEs have similar
optical structures - their radial profiles are exponential, at least at
large radii \cite{bmcm86,cb87,pt96,tmt97,mmhs97}.  Naturally this leads
to speculation: are there evolutionary connections between these
different morphologies?  One commonly accepted evolutionary path was
expounded by Davies \&\ Phillips \cite{dp88}.  In it, the initial
morphology is that of a dI.  If its ISM manages to concentrate at
the center of the galaxy a tremendous occurs starburst occurs resulting in a
BCD morphology.  Such starbursts are known to be capable of powering a
powerful galactic wind (e.g.\ \cite{ds86,mhws95,mfdc92}).  If the wind
is strong enough all of the ISM is expelled resulting in a dE
morphology.  If some ISM remains, the system fades back into a dI, and
undergoes a few more dI $\Leftrightarrow$ BCD transitions before
eventually expelling all of its ISM to become a dE.

Here I will address the validity of this scenario by piecing together
some work that I have been involved in
\cite{mmhs97,mmh98,mfdc92,mmc94,mcbf96,msk98} (citation implicit
throughout this review), as well as that of other researchers, that
answers some smaller questions.  What kind of dwarf galaxies host
starbursts?  What is the effect of starbursts on the ISM of dwarf
galaxies?  In addition I will consider another issue - dark matter (DM).
This is the dominant form of mass in dwarf galaxies.  Are the
differences in dwarf galaxy morphology related to DM content?  How does
DM effect the evolution of dwarf galaxies?

This review is heavily weighted towards gas rich dwarfs and in
particular BCDs.  In the Davies and Phillips scenario they are the
active link between dI and dE galaxies, and hence should provide the
best clues to deciphering dwarf galaxy evolution.  The paper is laid out
as follows: \S\ref{s:opt} compares the optical structure of dIs and
BCDs; \S\ref{s:rad} details the \HI\ structure and dynamics of two BCDs:
NGC~1705 ($D = 6.2$ Mpc) and NGC~2915 ($D = 3.1$ Mpc), and compares them
to dI galaxies; and \S\ref{s:syn} synthesizes the optical and radio
results to form a new scenario where DM plays a dominant role in
determining the morphology of gas rich dwarfs.  Throughout this review I
adopt distances based on $H_0 = 75\, {\rm km\, s^{-1}\, Mpc^{-1}}$.

\section{The optical structure and classification of gas rich dwarfs\label{s:opt}}

\begin{figure}
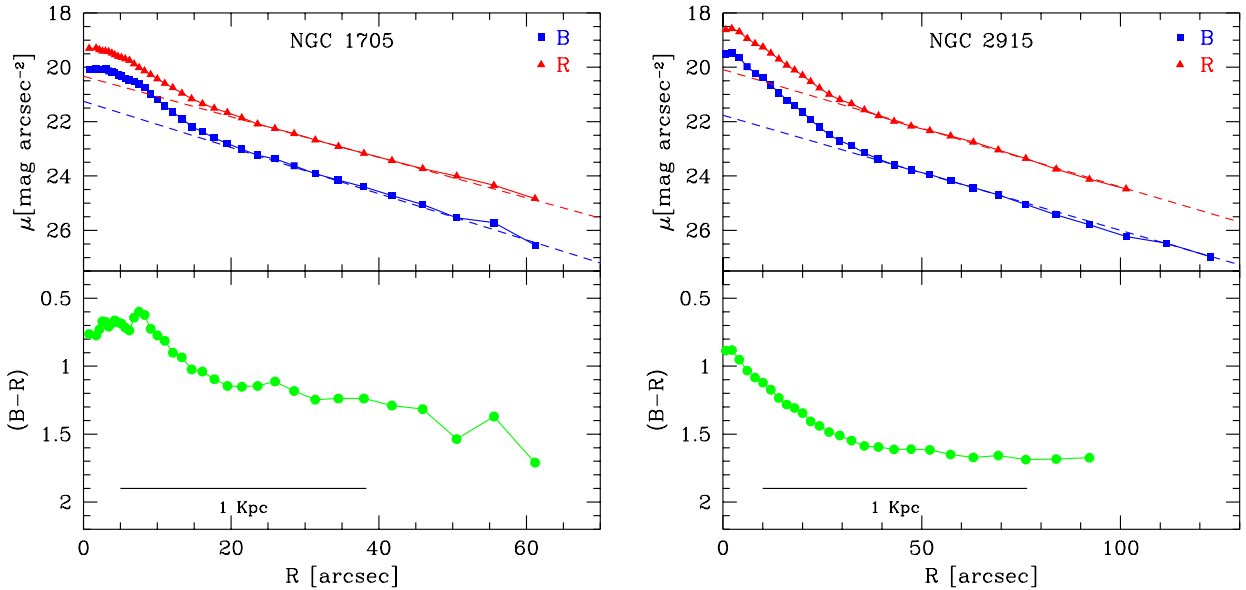

\centerline{\hbox{\psfig{figure=surf1705.cps,width=8.5cm}\psfig{figure=surf2915.cps,width=8.5cm}}}
\caption{Surface brightness profiles in $B$ and $R$, with corresponding
$(B-R)$ color profiles of NGC~1705 \cite{mfdc92} and NGC~2915
\cite{mmc94}.  The dashed lines show exponential
fits to the outer portions of the $\mu$ profiles.  Note: the optical
profiles of NGC~1705 exclude the light of the bright cluster NGC1705-1.
\label{f:oprof}}
\end{figure}

Figure~\ref{f:oprof} illustrates typical radial surface brightness
($\mu$) and color profiles of BCDs using NGC~1705 and NGC~2915 as
examples.  The exponential nature of the outer profiles is clearly
demonstrated by the linear fits in the $\mu$ versus radius $R$ plane.
While in dwarf galaxies it is not clear that the stars responsible for
this light are rotationally supported (in the lowest luminosity dEs they
are not \cite{mowfk91}), I will call this structure the {\em disk\/} for
brevity's sake and because both dIs and BCDs contain rotationally
supported \HI\ (\S\ref{s:rad}).  The central region {\em usually\/}
displays relatively blue emission in excess of the extrapolated disk. I
will call this structure the {\em starburst\/} because it is responsible
for the starburst characteristics of BCDs, as is clear from the
following evidence.  Firstly, \Halpha\ imaging shows that the most
intense \HII\ emission is confined to this region in BCDs.  Secondly,
HST \cite{cmbgkls97,m95} and ground based \cite{as85,cp89,mfdc92,mmc94}
imaging reveal numerous young blue star clusters in these structures.
Finally, their blue broad band colors indicate they must be due to young
stellar populations: they have ages of $\sim 10$ Myr if instantaneous
burst models are adopted or $\sim 100$ Myr if constant star formation
rate models are adopted.  The strong \Halpha\ fluxes from the cores are
more consistent with the constant star formation rate models (because
$\sim 10$ Myr old instantaneous bursts are no longer ionizing).  In
either case this is much less than the Hubble time, thus confirming
their starburst nature.  In comparison, the colors of the disk are
typically like those of stellar populations forming continuously over a
Hubble time (i.e.\ like dI galaxies, cf. \cite{hg85,pt96}), or a bit
redder suggesting an inactive population.  Hence, the disk is best
attributed to the progenitor, or {\em host\/}, galaxy.

Surface brightness profile fitting provides a means to determine both
the relative strength of the starburst, and the structural properties of
the disk.  The method is simple.  The outer portions of the profile are
fitted with an exponential. This yields the extrapolated central
surface brightness $\mu_0$ (which is corrected for inclination), and
scale length $\alpha^{-1}$ of the disk.  The burst and disk are
separated by assuming that the disk remains exponential all the way into
the center.

\begin{figure}
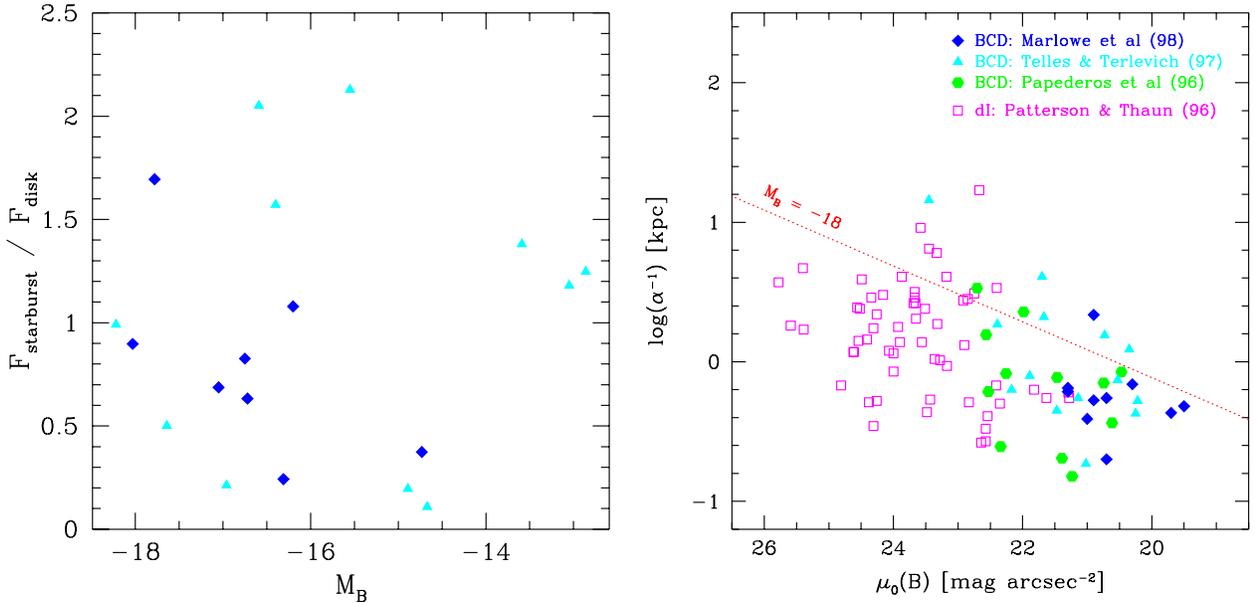

\centerline{\hbox{\psfig{figure=core.cps,width=8.5cm}\psfig{figure=struct.cps,width=8.5cm}}}
\caption{(Left) Ratio of $B$ band fluxes of starbursts to
host disks in two samples of BCDs (symbol correspondence in right
panel). \label{f:core}}
\end{figure}
\begin{figure}
\vspace{-0.5cm}
\caption{(Right) Structural parameters of the
exponential disks for a variety of samples of BCDs, and one dI
sample. \label{f:struct}}
\end{figure}

Figure~\ref{f:core} shows the ratio of $B$ band fluxes of BCD bursts
relative to their host disks.  The strongest starbursts are about twice
as bright as their hosts.  Hence, while starbursts can outshine the host
disk they are nevertheless modest\ltsim{1} mag enhancements
to the total $B$ flux of BCDs (cf.\ \cite{ss95}). Typical flux
enhancements are only a few tens of percent.  This figure does not
include upper limits to the burst/disk ratio: about 20\% of BCDs have
exponential profiles all the way into their cores (see below).  The mass
contribution of the starbursts is even smaller, typically\ltsim{5}\%.
These are not the $\sim 6$ mag starburst enhancements proposed to
explain the excess of faint blue galaxies at moderate redshifts
\cite{bf96}.

Figure~\ref{f:struct} compares the disk parameters $\mu_0$,
$\alpha^{-1}$ of both BCDs and dI galaxies.  I reemphasize that $\mu_0$
does not include the contribution of the starburst core.  While there is
some overlap, we see that BCD distribution is offset from that of dIs in
terms of both $\mu_0$ and $\alpha^{-1}$.  The offset in $\alpha^{-1}$ is
in the sense that BCDs tend to be smaller than dIs.  However this
difference may be due to the (loose) selection criterion $M_B
\gtsim -18$ mag typically applied to dwarf galaxy samples: large scale
length, high surface brightness galaxies are not dwarfs.  More striking
is the difference in $\mu_0$ (cf. \cite{ns97}). Typically $\mu_0$ is 2.5
mag arcsec$^{-2}$ more intense in BCDs than in dIs.  Structurally BCD
disks are very different from those of dIs.

The absence of BCDs on the left half of Fig.~\ref{f:struct} is puzzling.
Does this mean that dI galaxies do not experience starbursts?
Examination of the $\mu$ and color profiles of Patterson \&\ Thuan
\cite{pt96} reveal several dIs (e.g.\ UGC~5706, UGC~7636) with $\mu$
profiles of like those in Fig.~\ref{f:oprof}, that is containing a blue
central ``high'' surface brightness excess relative to the exponential
disk.  This is structural evidence for starbursts in dIs.  The episodic
star-forming nature of dIs is best demonstrated using color-magnitude
diagrams of the nearest ones (e.g.\ \cite{gtdschsm98}). However the
observed central surface brightness of the bursting dIs {\em including
the light of this central excess} is typically $\mu(B) \gtsim 22\, {\rm
mag\, arcsec^{-1}}$, much fainter than the central regions of BCDs.
While dIs do experience short duration bursts of star formation, they
are pathetic and not usually recognized as starbursts because they are
not {\em intense\/} enough.

On a similar note, about 20\%\ of BCDs (data from
\cite{mmhs97,pltf96,tt97}) have $\mu$ profiles that are nearly
exponential and have fairly flat color profiles (examples include Haro
14 \cite{mmhs97} and UM483 \cite{tt97}) - that is with no structural
evidence for a starburst.  The colors of these galaxies are consistent
with fairly long duration ($\gtsim 1$ Gyr) star formation. These results
highlight the importance of absolute surface brightness in morphological
classification.  For example, in the Virgo Cluster Atlas \cite{sbt85}
galaxies are recognized as dwarfs by their {\em low\/} surface
brightness.  Similarly BCDs are recognized by their {\em high\/} surface
brightness.  This is largely an unstated, perhaps even unconscious,
selection criteria.  There are clearly low surface brightness dwarfs
(e.g. GR8) that meet the usual luminosity, emission line, and size
criteria for BCD classification \cite{tm81} but are invariably
classified as dIs. From Fig.~\ref{f:struct} the BCD/dI dividing line is
$\mu_0(B) \approx 22\, {\rm mag\, arcsec^{-2}}$.  Clearly the presence
of a starburst makes it more likely that a dwarf will be classified as a
BCD.  However, such a classification does not guarantee the presence of
a strong burst.  Nor does the dI classification exclude galaxies
experiencing strong bursts relative to the host disk brightness.

\section{\HI\ structure and dynamics of BCDs\label{s:rad}}

Radio array observations of the \HI\ structure and dynamics of BCDs can
tell us much about star formation feedback (e.g.\ galactic winds) and
allow the distribution of mass (including DM) to be determined.
Compared to dIs there are not that many \HI\ imaging studies of BCDs; in
part because it is harder to find BCDs that are resolvable with radio
arrays, due to their small numbers and usual compact angular sizes.  My
collaborators and I therefore decided to address these issues with
Australia Telescope Compact Array observations of NGC~1705 and NGC~2915,
two of the nearer BCDs.  The resultant composite \HI\ and optical images
are shown in the color section of this volume.  Although both galaxies
have some kinematic irregularities their dominant structures are
extended rotating disks which are strongly centrally peaked.  These are
typical properties of BCDs imaged in HI \cite{s98,tbps94,v98}.  The disk
of NGC~2915 is so extended that it has the \HI\ appearance of a late type
barred spiral.  Similar galaxies include  IC~10 and NGC~4449 \cite{w98}.

\subsection{Starburst -- ISM feedback}

Both NGC~1705 and NGC~2915 show evidence of star formation churning up
the neutral ISM.  In NGC~2915, kinks and enhancements in the velocity
dispersion map correspond well to \Halpha\ bubbles and peculiar knots
associated with recent star formation.  However it does not appear that
\HI\ is being ejected from the system.  In the center of the galaxy,
where star formation is the most vigorous, $\sigma_{\rm HI} \approx 40$
\kms\ which is the same as the one dimensional velocity dispersion
derived for the DM particles (see Fig.~\ref{f:bcdrc} below).  Hence,
star formation appears to be maintaining the central \HI\ in virial
equilibrium with the DM halo.  This suggests that DM plays a role in the
feedback process: if the starburst energizes \HI\ to have $\sigma_{\rm
HI}$ much larger than the halo velocity dispersion, then neutral ISM is
thrown into the halo (or beyond) and star formation shuts down.

Jutting out obliquely from NGC~1705's edge-on \HI\ disk out to $R
\approx 4.5$ kpc, is a gaseous spur containing 8\%\ of the \HI\ flux
(see color plate).  Its position angle is similar to that of the
optical outflow, hence this may be the one-sided blow out of the
prominent \Halpha\ wind.  Its outflow is inferred to be
predominantly transverse, as is the \Halpha\ outflow, hence the
expansion velocity $V_{\rm exp}$, or similarly the expansion timescale
$t_{\rm exp}$, is uncertain.  The \Halpha\ outflow has $t_{\rm exp}
\approx 10$ Myr, like the age of the most prominent star cluster
NGC1705-1.  Adopting this $t_{\rm exp}$ for the \HI\ outflow yields
$V_{\rm exp} \approx 300$ \kms, an average mass loss rate of $\dot{\cal
M} \approx 1.7\, \Msun\, {\rm yr^{-1}}$, and total kinetic energy
$E_k({\rm gas}) \approx 1.5 \times 10^{55}\, {\rm erg}$.  If $t_{\rm
exp} \approx 100$ Myr, closer to the age of the continuosly star forming
population in the core of NGC~1705, then $V_{\rm exp} \approx 30$ \kms,
$\dot{\cal M} \approx 0.17\, \Msun\, {\rm yr^{-1}}$, and $E_k({\rm gas})
\approx 1.5 \times 10^{53}\, {\rm erg}$.  In comparison, the current
star formation rate is $\dot{\cal M}_\star \approx 0.13\, \Msun\, {\rm
yr^{-1}}$. Mass loss is at least competitive with star formation in
regulating the gas content of NGC~1705.  

Can the energetics of this gas be accounted for by NGC~1705's stellar
populations?  Using the solar metallicity Salpeter IMF (mass range 1 to
100 \Msun) population models of Leitherer \&\ Heckman \cite{lh95} I
estimate the mechanical energy output (from stellar winds, supernovae)
of the young stellar populations in NGC~1705 integrated over $t_{\rm
exp} = 10\, (100)$ Myr to be $E_k(\star) \approx 2.3\, (11.3) \times
10^{55}$ erg. The uncertainties on both $E_k({\rm gas})$ and
$E_k(\star)$ are probably a factor of few.  Hence, within the
considerable uncertainties, the energetics of the spur are are
consistent with it being a starburst driven wind.

The fate of NGC~1705's ejected gas depends critically on $V_{\rm exp}$.
If it is as high as 300 \kms, then even the large amounts of DM in the
system do not gravitationally bind it.  If $V_{\rm exp} \approx 30$
Myr, some gas will be retained.  At least 8\%\ of the neutral ISM has
been ejected into the halo of NGC~1705 if not out of the system
entirely.  Hence we are witnessing a significant mass ejection event.
Nevertheless, even in NGC~1705, a BCD with one of the most spectacular
\Halpha\ outflows, the majority of the ISM is retained in a disk.  Even
this starburst is incapable of totally blowing away the ISM.

\subsection{Mass distributions}

\begin{figure}
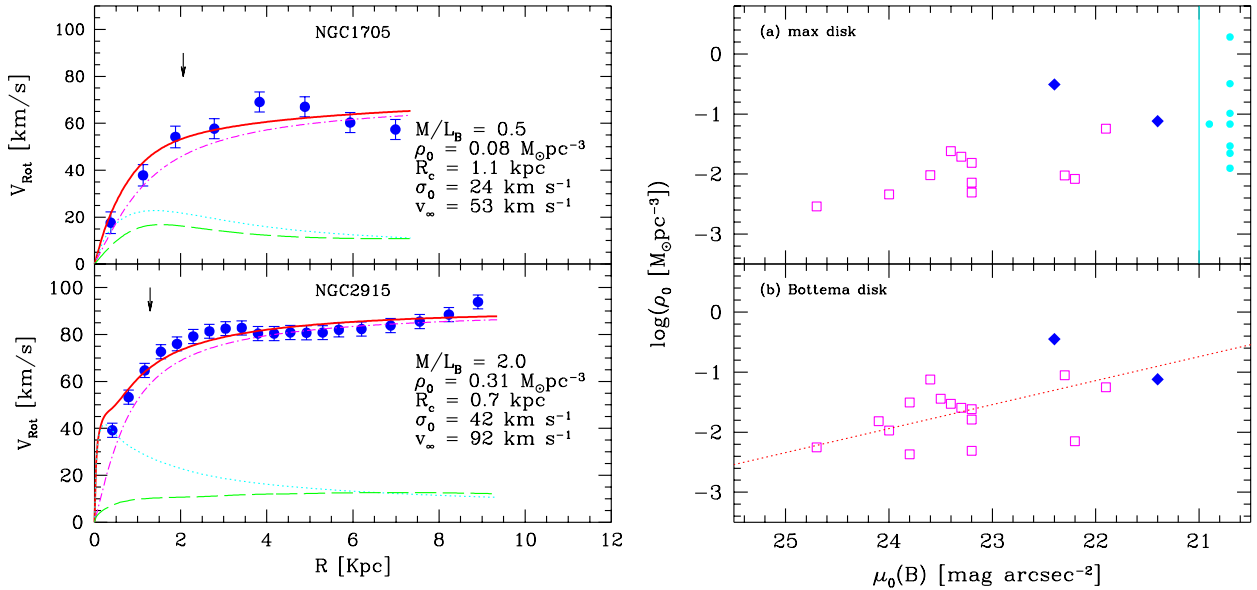

\centerline{\hbox{\psfig{figure=bcdrc.cps,width=8.5cm}\psfig{figure=rho0.cps,width=8.5cm}}}
\caption{(Left) Rotation curves of NGC~1705 \cite{msk98} and NGC~2915
\cite{mcbf96} with mass model fits.  The thick solid lines shows the
full models, the dotted, dashed, and dot-dashed lines show the
contributions of the stars, \HI\ disk, and DM halo respectively.  The
model parameters are listed in each panel.  The vertical arrows indicate the
optical Holmberg radius of each galaxy.  \label{f:bcdrc}}
\end{figure}
\begin{figure}
\vspace{-0.5cm}
\caption{(Right) DM halo central density $\rho_0$ plotted against disk
central surface brightness.  Open squares are from de Blok \&\ McGaugh
\cite{dm97}, while diamonds represent NGC~1705 and NGC~2915.  The top
panel shows the results for maximimum disk model fits, while the bottom
panel shows Bottema disk fits.  The circles on the right side of the top
panel mark crude estimates of $\rho_0$ in 12 BCDs with published and
unpublished RCs.  The dotted line, at bottom,  is a fit to the Bottema disk
results with the relationship $\log(\rho_0) = 0.4 \log(\mu_0) + {\rm
Constant}$.  \label{f:rho0}}
\end{figure}

Figure~\ref{f:bcdrc} shows the derived rotation curves (RCs) of NGC~1705
and NGC~2915 with mass model fits to them.  These are the first two BCDs
that have been mass modelled.  The models consist of three components to
the mass distribution: (1) the stellar distribution which is given by
the projected luminosity profile (Fig \ref{f:oprof}) scaled by \Mlbstar\
-- the mass to light ratio of the stars; (2) the neutral ISM
distribution which is set by the \HI\ profile scaled by a constant 1.33
to account for the Helium contribution (i.e.\ no free parameters); and
(3) a dark matter halo.  This halo is taken to be a
pseudo-isothermal sphere with a density distribution given by
\begin{equation} 
\rho = \frac{\rho_0}{1 + (R/R_c)^2} \label{e:mm}
\end{equation}
where the free parameters are the central density $\rho_0$ and the core
radius $R_c$. From these, the rotational velocity at
large $R$, $V_{\infty}$, and halo velocity dispersion $\sigma_0$ are
given by \cite{lsv90}:
\begin{equation} 
V_{\infty}^2 = 4 \pi G \rho_0 R_c^2 = 4.9 \sigma_0^2. 
\end{equation}

The models shown are maximum disk mass models, where \Mlbstar\ is set by
the first points in the RC, and then held fixed.  Maximum disk models
are by nature minimum halo models, hence $\rho_0$ is a lower limit.  We
also made minimum disk models ($\Mlbstar = 0$) and ``Bottema Disk''
models where \Mlbstar\ is set by the color of the optical disk
\cite{dm97}.

There are a few things to note about Fig.~\ref{f:bcdrc}.  First, the
form of the RCs is like that seen in normal disk galaxies, consisting of
a rising inner portion and a more or less flat RC thereafter. Second,
the optical extent of the galaxy, as marked by the arrows is contained
within the rising portion of the RC.  In fact the optical extent marks
very well the RC knee (or equivalently $R_c$).  The optical portions of
dIs are also commonly contained within the rising portion of their RCs
\cite{s87}.  This is where the rotation is almost solid body, and thus
shear is minimized, enhancing the ability of clouds to form stars
\cite{kcy93}.  Thirdly, in both cases DM dominates the mass
distribution, even within the optical radius of the galaxy.  In
comparison, the stellar component has a mass equal to or {\em less
than\/} the neutral gas disk.

Overall, the global dynamics of BCDs appear to be similar to dIs: they
are dominated by rotating disks with normal looking RCs.  A distinction
between the two types is seen when the DM halo densities $\rho_0$ are
compared, as shown in Fig.~\ref{f:rho0}.  Central densities found by
maximum disk and Bottema disk fits are shown in separate panels.  The
comparison sample is taken from de Blok \&\ McGaugh \cite{dm97}, and
includes only galaxies with $M_B > -18$ mag.  It is comprised mostly of
dIs (left side of panel), but also includes three low luminosity spirals
having $\mu_0(B) \approx 22\, {\rm mag\, arcsec^{-2}}$.  This comparison
shows that NGC~1705 and NGC~2915 have two of the highest $\rho_0$
measurements of any dwarf galaxies.  In order to check that these
galaxies are typical, I crudely estimated $\rho_0$ from the central
velocity gradient for 12 BCDs with published or unpublished RCs, and
plotted them as circles at arbitrary $\mu_0$ in the top panel of
Fig.~\ref{f:rho0}.  These estimates are upper limits, since the
contribution of the baryonic components to the velocity gradients have
not been removed.  Nevertheless, the comparison indicates that NGC~1705
and NGC~2915 have normal $\rho_0$ for BCDs.  Figure~\ref{f:rho0} shows a
weak but noticeable correlation between $\log(\rho_0)$ and $\mu_0(B)$,
with higher surface brightness disks corresponding to higher $\rho_0$
halos.  This result holds for both maximum disk and Bottema disk
solutions, and was first noted (for Bottema disks) by de Blok \&\
McGaugh \cite{dm97}, who show that it extends to high luminosity disks.
Here we show that the correlation also includes BCDs.

\section{Evolutionary Connections\label{s:syn}}

The correlation in Fig.~\ref{f:rho0} can readily be explained by
considering the response of a self gravitating disk immersed in a DM
halo core of constant density $\rho_0$, i.e.\ where the rotation curve
is linearly rising.  Then, the angular frequency $\Omega$, $\rho_0$, and
dynamical time $t_{\rm dyn}$ are constant with radius and related by
\begin{equation}
\rho_0 = \frac{3 \Omega^2}{4 \pi G}\hspace{1cm} ,
 t_{\rm dyn} = \frac{\pi}{2\Omega}, \label{e:rhot}
\end{equation}
and Toomre's \cite{t64} disk stability parameter $Q$ is given by
\begin{equation}
Q = \frac{2\sigma_g\Omega}{\pi G \Sigma_g}, \label{e:q}
\end{equation}
where $\sigma_g$ is the gas velocity dispersion and $\Sigma_g$ is the
disk surface density.  Lower values of $Q$ indicate a higher degree of
self gravity.  In normal disk galaxies $Q$ typically remains radially
constant at $Q \approx 2$, and star formation correlates with
regions of somewhat lower $Q$ \cite{k89,f97}.  This suggests that star
formation regulates disk structure to retain constant $Q$. 
Kennicutt \cite{k98} finds that for normal disk galaxies the star
formation rate per area $\dot\Sigma_\star$ is given by
\begin{equation}
\dot\Sigma_\star \propto \frac{\Sigma_g}{t_{\rm dyn}}\label{e:slaw}.
\end{equation}
As noted in \S\ref{s:opt} the optical disks of dwarf galaxies
typically have colors indicating a long duration of star formation.  Let
us assume that these are normal disks and hence form stars
with the above star formation law and at a constant and universal $Q$.
Then equations~\ref{e:rhot}, \ref{e:q} and \ref{e:slaw} can be combined to
yield 
\begin{equation}
\dot\Sigma_\star \propto \sigma_g \rho_0.
\end{equation}
For a constant star formation rate population and a sufficiently blue
passband (e.g.\ $B$) the linearly surface brightness is proportional to
$\dot\Sigma_\star$.  Therefore, for DM dominated galaxies we
expect a simple correlation between linear surface brightness and
$\rho_0$.  This is consistent with the observed correlation, as shown by
the dotted line in Fig.~\ref{f:rho0}.  Meurer et al. \cite{mhlll97} find
a similar correlation between surface brightness and $\rho_0$ holds in
the center of starburst galaxies.  However for them it is normal
baryonic matter that dominates $\rho_0$ rather than DM.

In essence, the central mass density determines the equilibrium star
formation rate of the embedded disk.  Following the discussion in
\S\ref{s:opt}, the disk intensity largely determines whether a dwarf
galaxy is classified as a BCD or dI.  As noted earlier, the optical size
of dwarfs seems to be limited to $R_c$.  Hence, both DM halo parameters
are important in governing the morphology of gas rich dwarfs.  Bottema
disk mass model decompositions \cite{dm97} indicate that the two DM halo
parameters are uncorrelated.  This is contrary to cosmological
simulations showing DM halos forming a one parameter (mass) sequence
\cite{nfw97}.  The origins and implications of this contradiction
deserve further investigation.

Can there be evolution between dI and BCD classes?  While some evolution
in $\rho_0$ may be allowed, it is unlikely that there can be enough to
change a typical dI into a typical BCD. That would require a 2.5 mag
arcsec$^{-2}$ change in $\mu_0$ or equivalently a factor of ten change
in $\rho_0$.  The most obvious way to do that is to expand or contract
the halo by a factor of $e = 10^{1/3} \approx 2.2$ following a mass loss
(wind) or gain (accretion or spin down of an extended disk).  The mass
loss/gain fraction $f$ required to effect this change is given by
\begin{equation}
e = \frac{1}{1 - f}
\end{equation}
for homologous expansion/contraction that is slow relative to $t_{\rm
dyn}$ \cite{ya87}.  A factor of 10 change in $\rho_0$
thus requires a 55\%\ mass loss or gain.  The problem is that there
isn't that much {\em baryonic\/} mass in a dwarf galaxy.  To effect this
large of a change would require DM loss or gain.  This is not
feasible if DM is non-dissipative and feels only the force of gravity,
as is usually assumed.  I conclude that there is probably little dI
$\Leftrightarrow$ BCD evolution.  

If the ISM were removed from a dI or BCD, it could still plausibly
evolve into a dE (or dwarf spheroidal) galaxy.  However, as noted in
\S~\ref{s:rad} even in a dwarf galaxy undergoing a strong starburst with
a spectacular galactic wind (NGC~1705), the fractional loss of the ISM
is modest.  If this is typical, it would take on order of 10 bursts to
expel all the ISM from a BCD.  The bursts aren't strong enough, and the
ISM distributions are too flattened to allow a single burst expulsion of
the ISM \cite{dh94}.  Single bursts should have a more profound effect
on the chemical evolution of dwarf galaxies, since the hot metal
enriched ISM is preferentially lost in a starburst driven wind
\cite{dg90}.  This view is consistent with arguments by Skillman \&\
Bender \cite{sb95} against evolution between gas rich and gas poor
morphologies occurring commonly at the present epoch.  The demographics
of dwarf galaxy morphologies point to an environmental component to
their evolution.  Gas rich dwarfs are found in low density environments
where the frequency of external starburst triggers is low.  They survive
easily. The clock runs faster (more frequent triggers) in clusters, and
in addition ram pressure stripping would accelerate the removal of gas
from dwarfs, while tidal truncation of DM halos would assist galactic
wind losses. Hence it is not surprising that gas poor dEs are found more
often in clusters than the field.

\section{Conclusions}

We are now at a position to re-evaluate the Davies and Phillips
\cite{dp88} scenario for dwarf galaxy evolution.  The mechanisms they
invoke have clearly been verified.  Dwarf galaxies do experience
starbursts and these can expel some of the ISM.  Mass expulsion can
rival or surpass lock up into stars in regulating the gas content of
dwarfs.  However, the results of any single burst are not so severe.
Cataclysmic bursts are not common at the present epoch, and the milder
bursts that are observed may not be sufficient to change a galaxy's
morphological classification.  The morphology of a dwarf galaxy is
largely set by its enveloping dark halo, and is relatively impervious to
starbursts.

\acknowledgements{My collaborators on the papers relevant to this review
were Sylvie Beaulieu, Carla Cacciari, Claude Carignan, Michael Dopita,
Ken Freeman, Tim Heckman, Neil Killeen, Glen Mackie, Amanda Marlowe, and
Lister Staveley-Smith.  I thank them for all their efforts.  I am
grateful to Liese van Zee, Eric Wilcots, and Crystal Martin for
providing during the conference additional data that went into
Fig~\ref{f:rho0}, and to John Salzer for useful discussions. }

\end{document}